\documentclass{webofc}

\usepackage[varg]{txfonts}   
\usepackage{hyperref}
\usepackage{url}
\newcommand{\nn}{\nonumber}
\hypersetup{colorlinks=true,citecolor=blue,urlcolor=blue,linkcolor=blue}

\begin{document}

\title{Heavy top quark mass in the minimal universal seesaw model}
%
%

\author{\firstname{Albertus Hariwangsa} \lastname{Panuluh}\inst{1,2,4}\fnsep\thanks{\email{panuluh@usd.ac.id}} \and
        \firstname{Takuya} \lastname{Morozumi}\inst{1,3}\fnsep\thanks{\email{morozumi@hiroshima-u.ac.jp}} 
}

\institute{Physics Program, Graduate School of Advanced Science and Engineering, Hiroshima University,\\1-3-1 Kagamiyama, Higashi-Hiroshima 739-8526, Japan 
\and
           Department of Physics Education, Faculty of Teacher Training and Education,\\ Sanata Dharma University, Paingan, Maguwohardjo,
           Sleman, Yogyakarta 55282, Indonesia 
\and
           Core of Research for the Energetic Universe, Hiroshima University, 1-3-1 Kagamiyama,\\ Higashi-Hiroshima 739-8526, Hiroshima, Japan
\and
           Presenter at the International Workshop on Future Linear Colliders (LCWS 2024)           
          }

\abstract{We study the hierarchy between $M_T, v_L$, and $v_R$, the relevant energy scales of the Minimal Universal Seesaw Model (MUSM), where the two lightest quark families remain massless at tree level. We also predict the heavy top quark mass, $m_{t'}$. We do some numerical analysis using recent experimental data.  Our numerical analysis demonstrates that $M_T$ is sensitive to the values of the Yukawa couplings. The heavy top quark mass $(m_{t'})$ is predicted to be within the range from 1.4 TeV to 7.2 TeV. 
}
\vspace*{-40pt} 
\hfill 
\begin{minipage}{0.2\linewidth}
	\normalsize
	HUPD-2407 \\*[5pt]
\end{minipage}
\maketitle
\section{Introduction}
\label{intro}
One of the questions in the particle physics is the origin of the fermion mass hierarchy. Fermion masses vary widely, with no clear explanation for the large differences between them. This hierarchy is particularly evident in the quark sector. The discovery of the top quark by the CDF \cite{CDF} and D\O~\cite{DO} collaborations in 1995 further highlighted this issue. The observed top quark is much heavier than other quarks, which presents a puzzle in understanding the origin of mass differences in the quark sector. This mass hierarchy problem has motivated to extend the SM.  One extension of the SM that attempts to explain the fermion mass hierarchy is the Universal Seesaw Model (USM) prior top quark discovery in Refs. \cite{ber0,ber,raj,chang,dav,ber2,ber3} and after top quark discovery, e.g., in Refs. \cite{koide,satou,kiyo}. 

A more recent development of the USM, where the two lightest quark families are massless at tree level, has been proposed in Ref. \cite{morpanu}. This model, called the Minimal Universal Seesaw Model (MUSM), naturally explains the observed quark mass hierarchy in the third family. Some of its phenomenological implications, such as flavor-changing neutral current (FCNC) processes in the interactions between the Higgs and Z bosons with quarks, are discussed in detail. One of the new physics (NP) particles predicted by this model is the heavy top quark $(t')$, also referred to as a top quark partner in some references. Searches for this heavy top quark have been conducted, e.g., by the CMS \cite{CMS} and ATLAS \cite{ATLAS} collaborations and have been summarized by the Particle Data Group (PDG) \cite{pdg}. These results provide a lower bound for the mass of the heavy top quark. In this work, we will study the prediction of the heavy top quark mass in more detail in Ref. \cite{morpanu}.

\section{The model}
\label{sec-model}
In this section, we briefly introduce the model in Ref. \cite{morpanu}. The model based on $\mathrm{SU(3)_C}\times\mathrm{SU(2)_L \times SU(2)_R \times U(1)_{Y'}}$ gauge symmetry. The particle content of the model is SM particle with additional one up-type $(T)$ and one down-type $(B)$ vector-like quarks (VLQs) as partner of top and bottom quark, respectively. Additionally, there is $\mathrm{SU(2)_R}$ Higgs doublet $(\phi_R)$. The charge convention that used in the model is
\begin{equation}
Q = I^3_{L} + I^3_{R} + Y' \label{charge convention},
\end{equation}
where $Q, I^3_{L(R)}$, and $Y'$ are electromagnetic charge, left(right) weak-isospin, and $\mathrm{U(1)_{Y'}}$ hypercharge, respectively. The particle contents of the model are given in table \ref{table-1}.
\begin{table}[t!]
	\caption{Particle contents of the model, where $i\in \{1,2,3\}$ is the family index. The content of this table is taken from Table 1 in Ref.\cite{morpanu}.}
	\centering
	\label{table-1}       
	\begin{tabular}{lcccc}
		\hline
		Quark and Higgs fields & $\mathrm{SU(3)_C}$ & $\mathrm{SU(2)_L}$ &$\mathrm{SU(2)_R}$ & $\mathrm{U(1)_{Y'}}$ \\\hline
		$q_L^i=\left( \begin{array}{c}
			u_L^i	\\d_L^i
		\end{array}\right)$  	& \textbf{3} & \textbf{2} & \textbf{1} & 1/6 \\
		[0.8\normalbaselineskip]
		
		$q_R^i=\left( \begin{array}{c}
			u_R^i	\\d_R^i
		\end{array}\right)$	& \textbf{3} &  \textbf{1}& \textbf{2} & 1/6 \\[0.8\normalbaselineskip]
		
		$T_{L,R}$	& \textbf{3} & \textbf{1} &  \textbf{1}& 2/3 \\ [0.8\normalbaselineskip]
		
		$B_{L,R}$	& \textbf{3} &\textbf{1}  & \textbf{1} & $-1/3$ \\ [0.8\normalbaselineskip]
		
		$\phi_L=\left( \begin{array}{c}
			\chi_L^+	\\\chi_L^0
		\end{array}\right)$ 	& \textbf{1} &\textbf{2}  &\textbf{1}  &1/2  \\ [0.8\normalbaselineskip]
		
		$\phi_R=\left( \begin{array}{c}
			\chi_R^+	\\\chi_R^0
		\end{array}\right)$	&\textbf{1}  & \textbf{1} &\textbf{2}  &  1/2\\\hline
	\end{tabular}
\end{table}

The Lagrangian of Yukawa interactions and mass terms of VLQs is as follows \cite{morpanu}:
\begin{align}
	{\cal L}_{\text{yuk}}&= 
	- Y_{u_L}^3\overline{q_L^3}\tilde{\phi}_L T_R -Y_{u_R}^3 \overline{T_L} \tilde{\phi}_R^\dagger q_R^3 - \overline{q^i_{L}} y^i_{d_L} \phi_L B_R - \overline{B_L} y^{i\ast}_{d_R} \phi_R^\dagger q_R^i -h.c.\nn\\ &\hspace{10pt} - \overline{T_L}M_T T_R - \overline{B_L} M_B B_R -h.c., \label{quark Lagrangian}
\end{align} 
where the Yukawa couplings of the up-type quark ($Y_{u_L}^3$ and $Y_{u_R}^3$)  are real positive numbers in a specific weak-basis while the Yukawa couplings of the down-type quark ($ y^i_{d_L}$ and $y^{i\ast}_{d_R}$) are general complex vectors\footnote[1]{The details are given in Appendix A of Ref. \cite{morpanu}.}. The charge conjugation of Higgs fields is defined as \linebreak$\tilde{\phi}_{L(R)}=i\tau^2 \phi^\ast_{L(R)}$ where $\tau^a$ with $a \in \{1,2,3\}$ is the Pauli matrix. In the second line of Eq.(\ref{quark Lagrangian}), $M_T$ and $M_B$ are the VLQ mass parameters that were taken as real numbers.

The symmetry of the model is spontaneously broken in two stages. First, $ \mathrm{SU(2)_R \times U(1)_{Y'}} \rightarrow\mathrm{U(1)_Y}$ when the neutral component of the $\mathrm{SU(2)_R}$ Higgs doublet acquires non-zero vacuum expectation value (vev) $v_R$. After this first stage of symmetry breaking, the symmetry of the model reduces to the SM gauge symmetry, $\mathrm{SU(3)_C \times SU(2)_L \times U(1)_{Y}}$. In the second stage, $\mathrm{SU(2)_L \times U(1)_{Y}} \rightarrow\mathrm{U(1)_{\text{em}}}$ when the neutral component of the $\mathrm{SU(2)_L}$ Higgs doublet acquires non-zero vev $v_L$. The definition of the vevs are given as follows:
\begin{equation}
	\left\langle \phi_R \right\rangle = \frac{1}{\sqrt{2}} \left( \begin{array}{c}
		0	\\ v_R
	\end{array}\right), \qquad 	\left\langle \phi_L \right\rangle = \frac{1}{\sqrt{2}} \left( \begin{array}{c}
	0	\\ v_L
	\end{array}\right)
\end{equation}  
and satisfy $v_R \gg v_L$. One can follow the derivation in Ref. \cite{morpanu} and obtain the exact mass eigenvalue of top quark $(t)$ and heavy top quark $(t')$ as follows\footnote[2]{This exact mass eigenvalue was first introduced in Ref.~\cite{eftproc}. Details on the diagonalization of the quark mass matrix can be found in Appendix C of Ref. \cite{morpanu}. In this work, we focus solely on the top sector.}:
 \begin{align}
	m_{t} &= \frac{\sqrt{M_{T}^2 + (m_{u_R} + m_{u_L})^2}}{2} - \frac{\sqrt{M_{T}^2 + (m_{u_R} - m_{u_L})^2}}{2}  \label{top exact mass},\\
	m_{t'} &=  \frac{\sqrt{M_{T}^2 + (m_{u_R} + m_{u_L})^2}}{2}+ \frac{\sqrt{M_{T}^2 + (m_{u_R} - m_{u_L})^2}}{2}   \label{top partner exact mass},
\end{align} 
where,
\begin{equation}
	m_{u_R} = Y^3_{u_R} \frac{v_R}{\sqrt{2}}, \qquad 	m_{u_L} = Y^3_{u_L} \frac{v_L}{\sqrt{2}}.
\end{equation}

\section{Numerical analysis}
In this section, we discuss the hierarchy between $M_T, v_L$, and $v_R$ using the exact mass eigenvalues of the top quark and the heavy top quark given in Eqs. (\ref{top exact mass}) and (\ref{top partner exact mass}), respectively. We analyze the constraints on $M_T$ and $v_R$, while taking $v_L = 246.22$ GeV from Ref. \cite{pdg}. As explained in Ref. \cite{morpanu}, the lower bound constraint on $v_R$ is $v_R \gtrsim 10$ TeV\footnote[3]{This lower bound is derived from the lower limit on the mass of the $Z'$ boson,  which in turn constrains the mass of the $W_R$ boson mass. Assuming $g_R \simeq 1$, this gives the lower bound on $v_R$. For more details, see Ref. \cite{morpanu}.}.  Additionally, we use numerical input from Ref. \cite{pdg} where the top quark mass is $m_t = 172.57$ GeV and the lower bound for heavy top quark mass is $m_{t'} > 1310$ GeV. The Yukawa couplings $Y^3_{u_R}$ and $Y^3_{u_L}$ are free parameters. However, we take the upper limit of the Yukawa couplings are $Y^3_{u_R}, Y^3_{u_L} \leq 1$.

\begin{figure}[h]
	\centering
	\includegraphics[width=0.8\textwidth,clip]{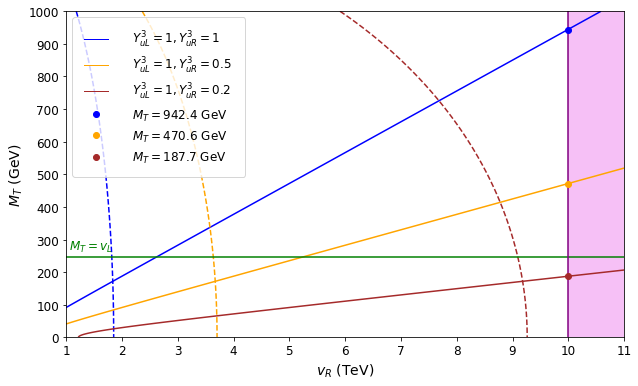}
	\caption{The variation of $M_T$ as a function of $v_R$, for different values of $Y^3_{u_R}$, with $Y^3_{u_L} = 1$. The vertical purple line represents the lower limit of $v_R = 10$ TeV. The solid blue, orange, and brown lines represent the constraints from the exact mass eigenvalue of the top quark ($m_t = 172.57$ GeV) for $Y^3_{u_R} = 1, 0.5, 0.2$, respectively. The corresponding dashed lines show the lower limit of the heavy top quark mass ($m_{t'} > 1310$ GeV) for each value of $Y^3_{u_R}$. The dots on the lines at $v_R = 10$ TeV indicate the respective $M_T$ values. The shaded pink region represents the allowed parameter space for $v_R \geq 10$ TeV.}
	\label{fig-1}       
\end{figure}

In figure~\ref{fig-1}, we explore the dependence of $M_T$ on $v_R$ for varying values of the $Y^3_{u_R}$ while keeping $Y^3_{u_L}=1$. The plot illustrates that as $Y^3_{u_R}$ decreases, the corresponding $M_T$ values also decrease for a given $v_R$. However, the variation of $Y^3_{u_L}$ is more stringently limited. It must satisfy $Y^3_{u_L} \geq  y_t^{\text{SM}}$ \cite{morpanu}, where $y_t^{\text{SM}}$ is the SM Yukawa coupling of the top quark, with a value of $y_t^{\text{SM}} \simeq 0.9912$.

Another important result is that two possible hierarchies emerge: $v_L < M_T < v_R$ or $M_T< v_L < v_R$, depending on the chosen parameters. After obtaining $M_T$, we compute the heavy top quark mass using Eq. (\ref{top partner exact mass}) and summarize the results in table~\ref{table-2}. The predicted heavy top quark mass is in the range of approximately 1.4 to 7.2 TeV, which could be tested in future linear collider experiments.

\begin{table}[t!]
	\caption{The heavy top quark mass for varying values of the $Y^3_{u_R}$ while keeping $Y^3_{u_L} =1$.}
	\centering
	\label{table-2}       
	\begin{tabular}{ccc}
		\hline
	$Y^3_{u_R}$& $M_T$ & $m_{t'}$ \\\hline
		1  	& $942.4$ GeV & $7.13$ TeV \\
		[0.8\normalbaselineskip]
		
		0.5	& $470.6$ GeV &  $3.57$ TeV \\[0.8\normalbaselineskip]
		
		0.2	& $187.7$ GeV & $1.43$ TeV \\ \hline
	\end{tabular}
\end{table}

\section{Conclusion}  
We have investigated the hierarchy between $M_T, v_L$, and $v_R$, and predicted the heavy top quark mass within the Minimal Universal Seesaw Model (MUSM), where the two lightest quark families remain massless at tree level. Our numerical analysis demonstrates that $M_T$ is highly sensitive to the values of the Yukawa couplings. Most importantly, the model predicts a heavy top quark mass in the range of approximately 1.4 to 7.2 TeV. This mass range represents a significant prediction that could be tested in future collider experiments, such as those planned for future linear colliders. 

\section{Acknowledgements}
We would like to thank the organizers of LCWS2024 for the opportunity to present our work at the conference.

%
%
%

\end{document}